\documentclass{aa}  
\include{journals}
\usepackage{rotating}
\usepackage{multirow} 
\usepackage{graphicx}
\usepackage{subfigure}
\usepackage{natbib}
\bibpunct{(}{)}{;}{a}{}{,} 
\usepackage{txfonts}
\newcommand{\rosat}{\textsc{rosat}}
\newcommand{\ace}{\textsc{ace}}

\newcommand{\stereo}{\textsc{stereo}}
\newcommand{\stereoa}{\textsc{stereo a}}
\newcommand{\plastic}{\textsc{plastic}}
\newcommand{\chandra}{{\it Chandra}}
\newcommand{\swift}{{\it Swift}}

\newcommand{\nen}{\ion{Ne}{ix}}
\newcommand{\net}{\ion{Ne}{x}}

\newcommand{\oxye}{\ion{O}{viii}}

\newcommand{\uvotw}{uvw1}
\newcommand{\vband}{v-band}

\def\opluss{O$^{7+}$}

\def\fluxseen{$4.3\pm 1.3\,\times\,10^{-13}$}

\def\ergsfu{$\mathrm{ergs}\,\mathrm{cm}^{-2}\,\mathrm{s}^{-1}$}
\def\uvfu{$\mathrm{W}\,\mathrm{m}^{-2}\,\mathrm{sr}^{-1}$}

\def\cts{$\mathrm{ct}\,\,\mathrm{s}^{-1}$}
\def\ctsa{$\mathrm{ct}\,\,\mathrm{s}^{-1}\,\mathrm{arcmin}^{-2}$}
\def\bone{0.3}
\def\btwo{1}
\def\deg{$^{\circ}$}
\def\mps{$\rm mol.\,s^{-1}$}
\def\moleps{$\times 10^{28}\,\mathrm{mol.}\,\mathrm{s}^{-1}$}
\def\molepsb{$\times 10^{29}\,\mathrm{mol.}\,\mathrm{s}^{-1}$}
\def\kms{$\rm km\,\,s^{-1}$}
\def\cmpers{$\rm cm^{-2}\,s^{-1}$}
\def\cms{$\rm cm^{2}$}
\def\cmd{$\rm cm^{-3}$}
\def\water{$\rm H_{2}O$}
\def\cdo{$\rm CO_{2}$}
\def\cmo{$\rm CO$}
\def\jand{2009-01-28}
\def\febd{2009-02-16}
\def\ofebd{2009-02-26}
\def\mard{2009-03-04}
\def\febf{2009-02-05}

\begin{document}
\title{Simultaneous Swift X-ray and UV views of comet C/2007 N3 (Lulin)}

   \subtitle{}

   \author{J.A. Carter\inst{1}
          \and
          D. Bodewits\inst{2}
          \and
          A.M. Read\inst{1}
          \and
          S. Immler\inst{2, 3}
          }

   \offprints{J.A. Carter}

   \institute{Department of Physics and Astronomy, University of
     Leicester, Leicester, LE1 1RH, UK\\
     \email{jac48@star.le.ac.uk}\\
     \email{amr30@star.le.ac.uk}\\
   \and
   Department of Astronomy, University of Maryland, College Park, MD 20742-2421, USA\\
     \email{dennis@astro.umd.edu}\\
   \and
   Astrophysics Science Division, Code 662, NASA Goddard Space Flight Center, Greenbelt, MD 20771, USA\\
     \email{stefan.m.immler@nasa.gov}\\
   }

   \date{Received 25 August 2011; Accepted 6 March 2012}

 
  \abstract 
{}
{We present an analysis of simultaneous X-Ray and UV observations of
  comet C/2007 N3 (Lulin) taken on three days between January 2009 and
  March 2009 using the \swift\ observatory.}
{For our X-ray observations, we used basic transforms to account for
  the movement of the comet to allow the combination of all available
  data to produce an exposure-corrected image. We fit a simple model
  to the extracted spectrum and measured an X-ray flux of
  \fluxseen\,\ergsfu\ in the \bone\ to \btwo\,keV band. In the UV, we
  acquired large-aperture photometry and used a coma model to derive
  water production rates given assumptions regarding the distribution
  of water and its dissociation into OH molecules about the comet's
  nucleus.}
{We compare and discuss the X-ray and UV morphology of the comet. We
  show that the peak of the cometary X-ray emission is offset sunward
  of the UV peak emission, assumed to be the nucleus, by approximately
  35,000\,km. The offset observed, the shape of X-ray emission and the
  decrease of the X-ray emission comet-side of the peak, suggested
  that the comet was indeed collisionally thick to charge exchange, as
  expected from our measurements of the comet's water production rate
  (6--8\,\moleps). The X-ray spectrum is consistent with solar wind
  charge exchange emission, and the comet most likely interacted with
  a solar wind depleted of very highly ionised oxygen. We show that
  the measured X-ray lightcurve can be very well explained by
  variations in the comet's gas production rates, the observing
  geometry and variations in the solar wind flux.}
{}

   \keywords{Comets: individual: Lulin - X-rays: general - Ultraviolet: general}
   \titlerunning{Comet Lulin observed by \swift}
   \authorrunning{}
   \maketitle
%

\section{Introduction}\label{sec:luli_intr}
Comets emit X-rays via the process of solar wind charge exchange
(SWCX). Gas in the coma of the comet donates one or more electrons
into an excited energy level of a highly-charged ion of the solar
wind. In the subsequent relaxation of the ion, a UV or X-ray photon is
emitted \citep{lisse1996, cravens1997, krasnopolsky1997}.

SWCX in comets probes states of the solar wind throughout the
heliosphere \citep{schwadron2000, kharchenko2001}. To date, over 20
comets have been observed in X-rays \citep{lisse1996, dennerl1997,
  lisse2004, krasnopolsky2006, bodewits2007}. This sample contains a
broad variety of comets, solar wind environments and observational
conditions and clearly demonstrates the diagnostics available from
cometary charge exchange emission. Several of those comets showed
interesting large scale structures in X-ray such as the jets in
2P/Encke \citep{lisse2005}, the Deep Impact triggered plume in
9P/Tempel 1 \citep{lisse2007}, and the disintegrating comets
73P/Schwassmann-Wachmann 3 and C/1999 S4 (LINEAR)
\citep{wolk2009}. These interpretations relied on ground-based
observations to initially identify those structures. Several attempts
have been made to explain the long-term variability of cometary X-ray
emission \citep{lisse1999, neugebauer2000, lisse2005, willingale2006,
  lisse2007} via relationships between the comet gas production rate,
the heliocentric distance, and the behaviour of the solar wind. This
paper however presents the first endeavour to directly constrain the
cometary gas production rates and relate this to the observed X-ray
variability by employing \swift's co-aligned instrument suite,
allowing us to make simultaneous X-ray and UV measurements from a
single spacecraft.

C/2007 N3 (Lulin) is a dynamically new comet that was discovered by
Lin Chi-Sheng and Ye Quanzhi at Lulin Observatory, Taiwan, in 2007. In
this paper we analyse data from comet Lulin obtained by the \swift\
satellite \citep{gehrels2004} on several days in early 2009 as part of
its Targets of Opportunity programme for non-gamma ray burst
targets. The days were chosen to sample emission from the comet at
various stages during its passage into and out of the inner solar
system, taking advantage of its high activity and its close proximity
to Earth. An initial analysis of these observations using data from
the \swift\ UV grism was discussed in \citet{btest2011}.

Comet Lulin moves in an orbit of low inclination of just 1.6$^\circ$
from the ecliptic, allowing us to link measured solar wind data from
the Solar Terrestrial Relations Observatory spacecraft (\stereo,
spacecraft A, part of the \plastic\ plasma package,
\citealt{blush2005}) to our observations. We expected that this comet
would sample the low-latitude, highly-charged solar wind and that the
intensity of this emission would reflect changes in the solar wind
flux, the quantity of neutral species in the comet's atmosphere and
the distance to the observer, similar to what has been seen in other
near-ecliptic comets (e.g. 2P/Encke \citep{lisse1999, lisse2005};
P/Tempel 1 \citep{lisse2007}).


The observations were taken around solar minimum, when the solar wind
can be simplified to the stratification of a slow, low-latitude wind
originating about the solar equator and a higher-latitude, faster
wind.  The heavy-ion freeze-in temperature at the solar corona
determines the ion composition of the out-flowing solar wind. In the
bimodal state outside solar maximum, the polar wind has a lower ion
freeze-in state than the wind at lower latitudes
\citep{geiss1995}. The nominal low-latitude solar wind flows radially
away from the Sun, yet the Sun is rotating, resulting in a Parker
spiral outflow \citep{parker1958}. However, this flow contains streams
of plasma with different radial velocities. Co-rotating Interaction
Regions (CIRs) occur when a faster stream piles up against the slower,
ambient plasma, resulting in an upstream compression region and a
downstream rarefaction region. CIRs are characterised by an increase
in ion density (with a thickness of $\sim$\,0.1\,AU) followed by a
decrease in this density, and a corresponding sharp increase in the
ion velocity \citep[and references therein]{gosling1999}. They are
found at low to mid heliospheric latitudes. Coronal Mass Ejections
(CMEs, or ICMEs when they are found within interplanetary space)
however, do not form part of the nominal co-rotating solar wind flow,
but are large clouds of fast moving plasma ejected from the Sun with
distinctive compositional signatures. Along with differences between
the fast and slow wind compositional fractionation as described above,
transient events such as ICMEs show marked abundance signatures that
can be used to identify plasma of this type. The presence of
highly-charged iron, elevated oxygen states along with enhanced
$\alpha$-to-proton ratios for example are indicators of ICME plasma
\citep{richardson2004, zurbuchen, zhao}.


This paper is organised as follows. In Section~\ref{sec:luli_data} we
discuss the data analysis steps taken for both the X-ray and UVOT data
sets. In Section~\ref{sec:luli_disc}, we discuss our background and
spectral analysis of the XRT data along with the X-ray radial extent
of the comet, the morphology of the X-ray and UV emission, the gas
production rates of the comet and the X-ray temporal variability. We
finish with our conclusions in Section~\ref{sec:luli_conc}.

\section{Data reduction}\label{sec:luli_data}

\subsection{X-ray telescope}\label{sec:luli_xrtd}

We use data from the X-ray telescope (XRT; \citealt{burrows2005}) in
photon counting mode. The XRT is sensitive over the energy range 0.2
to 10\,keV, has an effective area of $\sim$80\,\cms\ at 1\,keV
\citep{godet2009}, a spectral resolution of $\sim$130\,eV at
Si-K$\alpha$ \citep{pagani2011} and a sensitivity limit of
$2\,\times\,10^{-14}$\,\ergsfu. The primary source of the XRT
particle-induced background is protons incident on the focal plane,
swept up as \swift\ moves in its low-Earth orbit.

The 14 \swift\ XRT observations used in this analysis, taken on three
different days (\jand, \febd\ and \mard, 2009 UT) are listed in Table
1. These observations were at fixed pointings and \swift\ did not
track the comet. Ephemeris data for the comet were obtained from the
Horizons website\footnote{http://ssd.jpl.nasa.gov/horizons.cgi}. We
used the cleaned XRT event lists which have been filtered for periods
of high background by the standard \swift\ processing chain. The event
files had also been adapted for an improvement in the gain function of
the XRT (XRT instrument team, private communication). No bright X-ray
astronomical point sources were formally detected in the field-of-view
(FOV) for any of the pointings by the source detection algorithm
\textit{detect} of the HEASARC task
\textsc{ximage}\footnote{http://heasarc.nasa.gov/xanadu/ximage/ximage.html}
.

For every event file we transformed each individually recorded X-ray
photon in detector space, using the ephemeris information to account
for the comet's movement across the sky. We calculated the change in
Right Ascension and Declination of each event and calculated the new
detector coordinates accordingly. Every exposure was transformed so
that the nominal position of the comet at the exposure start was found
at the centre of the detector, and we accounted for the change in
Sun-direction by rotating each frame. We then stacked these event
files together to improve the signal to noise and adapted the exposure
keywords appropriately. Vignetted exposure maps were produced for each
observation. Each exposure map was stretched to account for the
movement of the comet and shifted and rotated as per the steps applied
to the event files. All exposure maps were stacked to provide a
combined exposure map to accompany the stacked event file.

\begin{figure}
  \centering
  \includegraphics[width=0.45\textwidth]{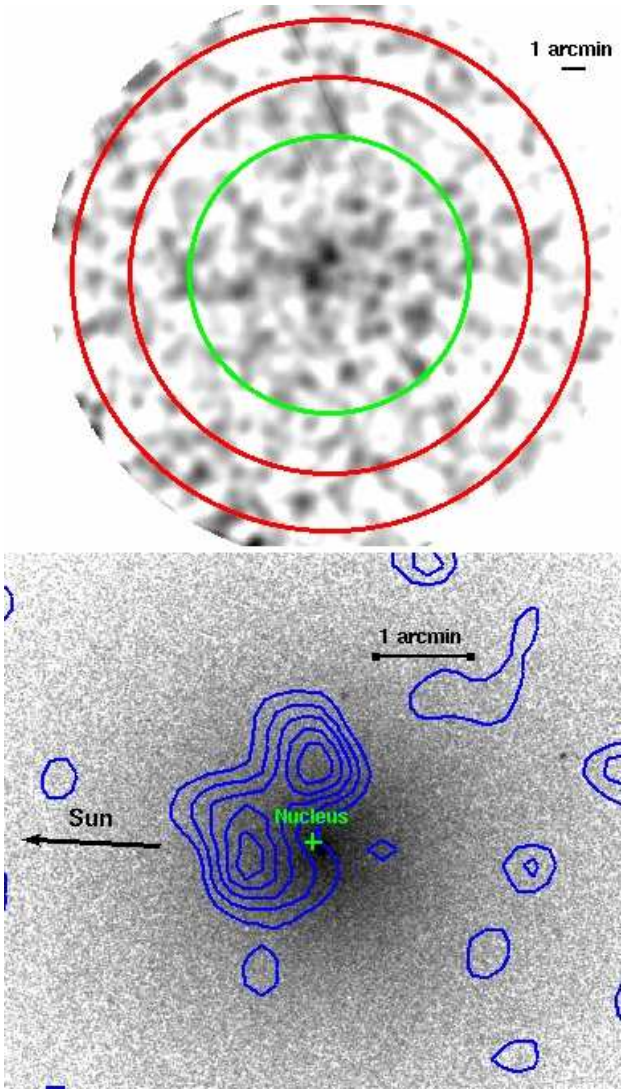}
  \caption{Upper panel: a smoothed \bone\ to \btwo\,keV exposure
    corrected image of the stacked XRT event list following the
    transformations described in Section~\ref{sec:luli_xrtd}. The
    circle (green) and annulus (red) refer to regions used for
    spectral analysis, see Section~\ref{sec:luli_xrts}. Lower panel:
    zoomed-in UV image of the comet, from the first UVW1 exposure of
    observation 31332 (see Table 1), taken with a
    pointing Right Ascension and Declination of (231.44\deg\ and
    -17.89\deg), with X-ray contours from the upper image
    overlaid. Contours start at 1.0\,$\times \, 10^{-3}$\,\ctsa\
    ($\sim$~6.5 sigma above the background), and increase in increments
    of 0.4\, $\times \, 10^{-3}$ \ctsa. A one arcminute bar is shown
    on each panel.}
  \label{fig:luli_imag}
\end{figure}

We created a smoothed image of the stacked events and divided this by
the stacked exposure map to produce an exposure-corrected image. As
there was no significant emission detected above 1\,keV, we study only
photons with E~$<$~1\,keV here. We restrict the lower limit of our
analysis to 0.3\,keV due to the uncertain calibration of the XRT below
this energy \citep{godet2009}. A \bone\ to \btwo\,keV
exposure-corrected image is shown in Figure~\ref{fig:luli_imag} (upper
panel), which has been rotated so that the Sun vector is as for the
first observation of the data set. This image has been smoothed
employing a Gaussian filter with FWHM 4.6 pixels (one pixel equals
2.36 arcseconds side length).

\begin{figure*}
  \centering
  \includegraphics[bb=125 50 460 750, clip]{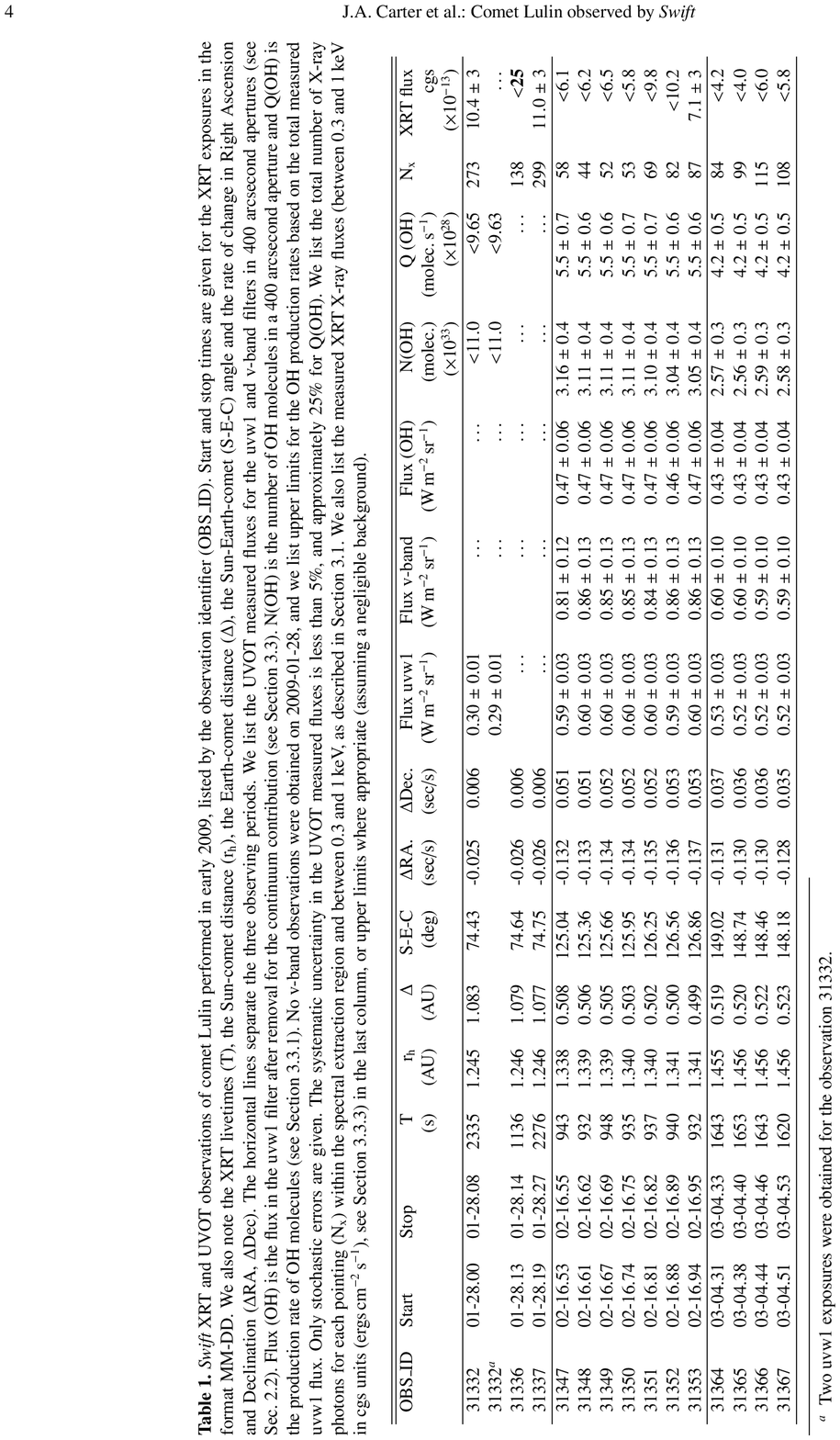}
  \label{tab:luli_obsn}
\end{figure*}

\subsection{UV-Optical telescope}\label{sec:luli_uvot}
Along with the X-ray observations, the \swift-Ultraviolet/Optical
Telescope (UVOT; \citealt{roming2005}) acquired 75 observations of
comet Lulin using its optical and UV broad-band filters. UVOT provides
a $17 \times 17$ arcminute FOV, with a plate scale of 1
arcsecond/pixel and a point spread function of 2.5 arcseconds FWHM
\citep{mason2004}. Additionally, on \jand\ the UV grism was used to
acquire low resolution ($\lambda/\delta\lambda = 100$) spectra of the
comet \citep{btest2011}. UVOT images of comet Lulin using the \uvotw\
($\lambda_c$ 2600\AA, FWHM 700\AA) and \vband\ ($\lambda_c$ 5468\AA,
FWHM 750\AA) filters are discussed later in
Section~\ref{sec:luli_morp}.

Water vapour and its fragment species are the most abundant volatiles
in cometary comae. Comet Lulin produced no more than 2-5\% \cdo\ and
$<$2\% \cmo\ \citep{btest2011, ootsubo2010}. Given that these
molecules have comparable charge exchange cross sections at solar wind
ion velocities as \water\ \citep{mawhorter2007}, the number of water
and hydroxyl molecules is a good proxy for the comet's contribution to
the variability of the X-ray lightcurve. We used \swift's \uvotw\ and
\vband\ filters to measure the number of OH molecules in the FOV, and
to estimate Lulin's water production rate during our observations.

We measured the comet's flux in large apertures of 400 arcseconds
radius to get excellent signal to noise, to sample a large fraction of
the coma, and to allow for comparison with the X-ray photometry
(Section~\ref{sec:luli_morp}). We obtained coma profiles by
azimuthally averaging the surface brightness. This was achieved by
converting each UVOT image into polar coordinates ($r, \phi$) and
finding the mean surface brightness at a given radial distance
($r$). By taking the azimuthal-averaged surface brightness for every
pixel at a given radial distance ($r$) from the optocenter (area of
brightest emission, that is not necessarily coincident with the
position of the nucleus), we constructed a 2-dimensional azimuthal
average image of the comet that was used to search for faint
structures.

Although this method assumes the large scale coma to be symmetric, it
enables the effective filtering out of stars and other background
objects. The coma fills the entire UVOT FOV, but by comparing the
signal at the edge of the detector with \swift\ calibration data we
estimated a background signal of 0.005 counts~s$^{-1}$~pixel$^{-1}$
(\uvotw) and 0.02 counts~s$^{-1}$~pixel$^{-1}$ (\vband), corresponding
to 1\% (\uvotw) and 13 \% of the total signal from the comet. At the
comet's optocenter, the count rates were high enough to result in some
coincidence loss (here approximately 7\% for \uvotw, and $\leq$20\%
for \vband) of the flux in a 5 arcseconds aperture around the
optocenter \citep{poole2008}. We decided not to correct our photometry
for this coincidence loss, as it only affected a small fraction of the
total flux in the 400 arcsecond aperture. Based on these
considerations we estimate stochastical errors to be 5\% (\uvotw) and
15\% (\vband). Absolute errors in the photometry are better than 3\%
\citep{breeveld2011}.

The measured fluxes in the filters relevant to this study (\uvotw\ and
\vband) are summarised in Table 1. Note that two UVOT exposures with
the same observation ID (31332) were obtained in the same \swift\
orbit, while in observations 31336 and 31337 no exposures with the
\uvotw\ or \vband\ filters were obtained.

\section{Discussion}\label{sec:luli_disc}

\subsection{Background and spectral analysis of the XRT
  data}\label{sec:luli_xrts}
\begin{figure}
  \centering
  \includegraphics[width=0.37\textwidth, angle=270, clip]{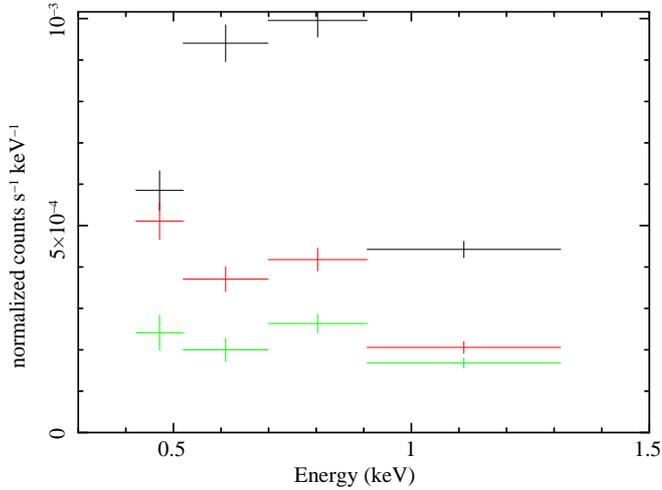}
  \caption{RASS backgrounds for the three sky positions of the comet
    (see text and Table 1), for \jand\ (black), \febd\ (red) and
    \mard\ (green).}
  \label{fig:luli_rass}
\end{figure}

Comets are extended sources, often larger than the FOV of the
observing instrumentation. Additionally, the X-ray background can show
considerable temporal and spatial variation. The background
subtraction is therefore non-trivial, and we compare two different
approaches here. First, we downloaded three \rosat\ all-sky survey
spectra (radius 1 degree) and the \rosat\ PSPCc response matrix from
the HEASARC background
tool\footnote{http://heasarc.gsfc.nasa.gov/cgi-bin/Tools/xraybg/xraybg.pl},
where any known so-called Long Term Enhancements, due to time-variable
SWCX (from exospheric or heliosphere emission,
e.g. \citealt{collier2005} or \citealt{koutroumpa2006}) have been
removed \citep{snowden1995}. The three pointings were chosen to
reflect the comet's position on the sky during the three observing
periods (see Table 1). We plot the three background
spectra (from OBS\_IDs 4973969, 4917081, 4814705) in
Figure~\ref{fig:luli_rass}. The spectra show a considerable amount of
variation. It was decided however, that creating three separate
spectra for the comet would be infeasible due to the poor statistics
that would be obtained.

\begin{figure}[h]
  \centering
  \includegraphics[width=0.475\textwidth, clip]{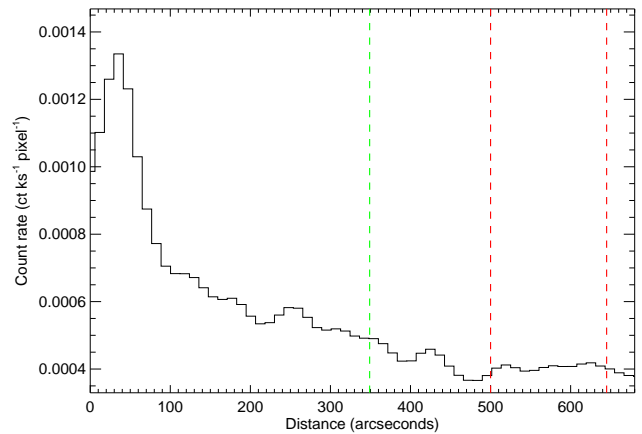}
  \caption{Radial profile of the \bone\ to \btwo\,keV combined XRT
    exposure-corrected image. The dashed vertical lines indicate the
    radius of the spectral (green) and the boundaries of the
    background extraction regions (red), chosen to maximise the signal
    to noise and to avoid subtracting signal coming from the comet.}
  \label{fig:luli_radp}
\end{figure}

The second method, which we deemed more reliable, concentrates instead
on a spectrum obtained from the combined spectrum from all the
available data for the comet. We show a radial profile of the exposure
corrected image in Figure~\ref{fig:luli_radp}. Distances are quoted in
arcseconds from the centre of the detector (which, following the
procedure described in Section~\ref{sec:luli_xrtd} is the centre of
the comet according to the ephemeris transformations that we had
applied to the X-ray data). The radial profile allowed us to constrain
the area in the detector plane that we considered to contain emission
from the comet by selecting a radius at which the profile drops to the
apparent background level. The peak of the radial profile is offset
from the origin of the distance axis. This offset is described in more
detail in Section~\ref{sec:luli_morx}. We extracted a background
spectrum from the stacked event file, using a centrally positioned
annulus with inner and outer radii of 500 and 645\,arcseconds. We then
extracted a spectrum using a centrally positioned circular extraction
region, with a radius of 349\,arcseconds chosen to maximise the signal
to noise. Both regions are shown overlaid on the exposure-corrected
image of the stacked events of Figure~\ref{fig:luli_imag} (upper
panel). We applied the appropriate instrument XRT response matrix and
effective area correction file (photon counting mode).

\begin{figure}[h]
  \centering
  \includegraphics[width=0.35\textwidth, angle=270, bb=75 -36 600 700, clip]{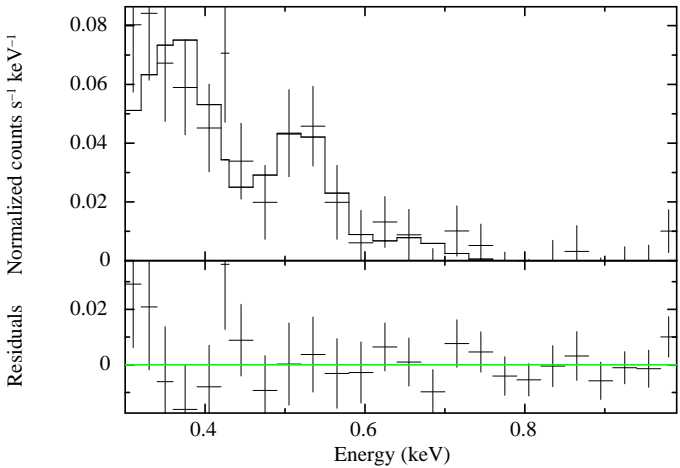}
  \caption{Upper panel: background subtracted spectrum from the
    combined comet Lulin XRT data set. The model fitted to the data,
    consisting of three zero-width Gaussian lines as described in the
    text, is shown by the solid line. Lower panel: residuals to the
    model fit.}
  \label{fig:luli_spec}
\end{figure}

The SWCX emission process is characterised by line emission in the
soft X-ray band. Using the
XSPEC\footnote{http://heasarc.gsfc.nasa.gov/docs/xanadu/xspec/index.html}
spectral fitting package we fit an un-absorbed SWCX-like simple model
to the background-corrected spectrum (shown in
Figure~\ref{fig:luli_spec}), incorporating three zero-width Gaussian
lines, letting the energies and the line normalisations vary. The
lines were found at 0.38\,keV, 0.52\,keV and 0.64\,keV, which we
attribute to a blend of helium- and hydrogen-like carbon and nitrogen
lines, nitrogen and oxygen lines\textbf{,} and oxygen
respectively. The reduced-Cash statistic (maximum likelihood-based
statistic for Poisson data, \citealt{cash1979}) was 1.3 for 63 degrees
of freedom with a not unreasonable goodness of fit of 35\%. The
goodness of fit should be approximately 50\% had the observed spectrum
been produced by the model. The un-absorbed flux between \bone\ and
\btwo\,keV was found to be \fluxseen\,\ergsfu\ (95\% confidence). We
also note the background-subtracted count rates in two energy bands of
interest: 4.1\,$\pm$\,1.1\,$\times 10^{-3}$ \cts (0.3 to 0.5\,keV) and
4.1\,$\pm$\,1.6\,$\times 10^{-3}$ \cts (0.5 to 0.7\,keV). We look into
the variation of flux with respect to time in
Section~\ref{sec:luli_xrlc}, in comparison to the behaviour of the
solar wind at the time of the observations.

Several sophisticated methods have been developed to analyse cometary
SWCX spectra \citep{beiersdorfer2003, kharchenko2003,
  krasnopolsky2006, bodewits2007}. We chose to apply a much simpler
model here due to the low-quality spectrum achieved, even after
stacking of the entire data set to improve the signal to noise. We do
not observe any emission above $\sim$0.8\,keV, in contrast to comets
within the \chandra\ survey \citep{bodewits2007}, which consistently
required higher-order transitions of \oxye\ plus two Ne lines at
0.90\,keV (\nen) and 1.024\,keV (\net). The \swift\ XRT has an
effective area of $\sim$\,65\,\cms\ at $\sim$\,0.8\,keV, much lower
than the $\sim$\,300\,\cms\ at the same energy of the \chandra\ ACIS
S3 chip used in the \citet{bodewits2007} survey. However, the \swift\
XRT effective area increases to a maximum of $\sim$\,120\,\cms\ at
about $\sim$~1.5\,keV (compared to $\sim$\,700\,\cms\ for the
\chandra\ ACIS S3 chip), implying that the lack of emission above this
energy is due to incoming solar wind compositional effects, which we
discuss more in Section~\ref{sec:luli_xrlc}. Although the poor
signal-to-noise ratio inhibits interpretation of the signal at higher
energies, the lack of such a signal may suggest that the comet
interacted with a solar wind with a charge state distribution
reflecting a very low freeze-in temperature. The spectral shape of the
model applied here most closely resembles that of comets 2P/Encke
\citep{lisse2005} and 73P/Schwassmann-Wachmann 3 \citep{wolk2009} of
the \chandra\ sample, which sampled the coolest wind of that
survey. We comment in more detail about the composition of the
interacting solar wind in Section~\ref{sec:luli_xrlc}.

\subsection{Morphology: UV and X-ray}\label{sec:luli_morp}
\subsubsection{Coma and tails}
\begin{figure*}
  \centering
  \includegraphics[width=0.99\textwidth, angle=0]{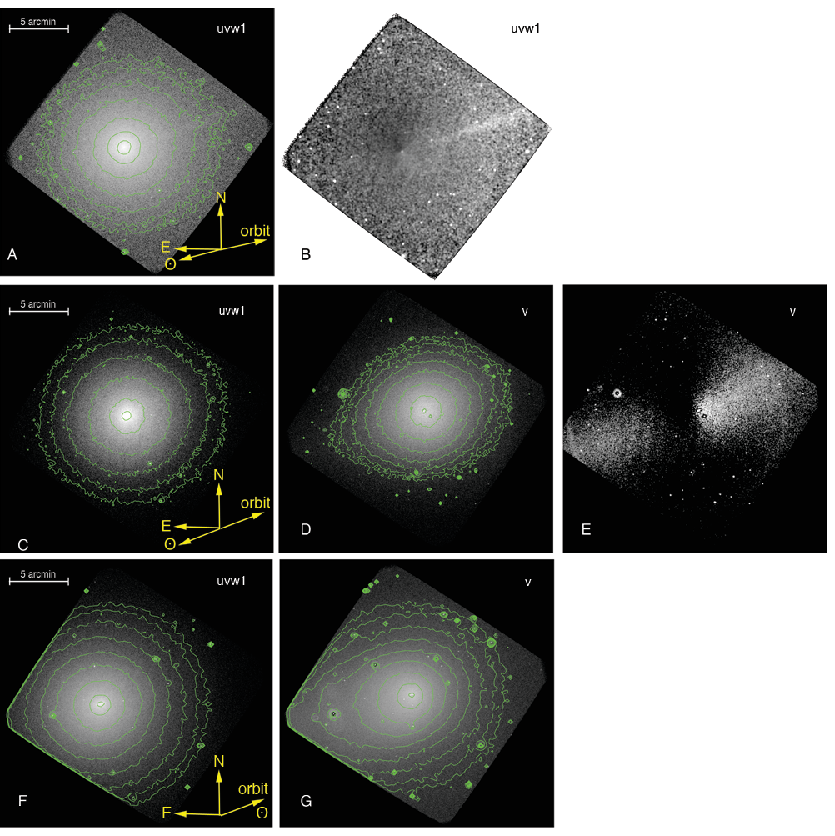}
  \caption{UVOT observations of Comet Lulin. {\bf A.} \jand,
    \uvotw. {\bf B:} image A divided by coma profile. {\bf C:} \febd,
    \uvotw. {\bf D:} \febd, \vband.  {\bf E:} image D with the coma
    profile subtracted. {\bf F:} \mard, \uvotw\ {\bf G:} \mard,
    \vband. All images are oriented in the same way (east left, north
    up) and have the same angular scale (UVOT's FOV is 17 $\times$ 17
    arcminute). Iso-intensity contours are shown on a logarithmic
    scale to enhance fainter features. In panel {\bf F} the solar
    direction vector and the orbital vector are coincident.}
  \label{fig:luli_uvot}
\end{figure*}

UVOT images for every observing day are shown in
Figure~\ref{fig:luli_uvot}. The left most column shows the comet in
the \uvotw\ band for each observing periods (panels A, C and F). In
this band the comet appears very symmetric around the nucleus,
confirming that most of the emission seen in the \uvotw\ band comes
from OH molecules (Section~\ref{sec:luli_gasp}). The comet's
morphology in the \vband\ is driven by the changing viewing
geometry. No \vband\ observations were obtained on \jand. On \febd\
the dust coma is elongated in the solar direction. Subtracting the
azimuthal profile as determined by the method described in
Section~\ref{sec:luli_uvot} (c.f. \citealt{schleicher2004}) reveals
both the dust tail (westward) and the anti-tail (eastward); the latter
consists of long lived dust emitted near perihelion (panel E). Lulin
moves in a retrograde orbit and Earth crossed the comet-Sun plane
around \ofebd. On \mard, the observing geometry therefore no longer
allows easy separation of particles along different trajectories,
resulting in a dust tail in the anti-solar direction (shown by the
elongated contours in panel G).

On \jand, a well defined, narrow tail can be seen to extend westwards
(best seen in panel B, where the image is divided by the azimuthal
profile). It is not visible in the \febd\ images (panel E). The nature
of this narrow tail was first discussed in \citet{btest2011}. In this
paper it was argued that from the geometry the narrow tail could be
either the dust tail or the ion tail (both clearly visible in wide
field amateur images obtained in this period), but that while the
\uvotw\ filter has a significant red-leak, it is not very sensitive to
the emission usually associated with the blue CO$^+$ comet tail system
(4000 -- 5000 \AA). It was concluded that it was more likely that the
narrow feature arose from continuum emission from dust or ice. No
asymmetric feature, however, is visible in the \uvotw\ and \vband\
azimuthal profile-subtracted images on \febd.

As we argued above, the UV filter transmission means that the narrow
tail is likely to be small-grained dust rather than ions. The
geometric conditions and the fact that the tail was detected in the UV
observations in January but not in February suggests that it is the
result of enhanced activity of these grains around perihelion. The
dust, released around perihelion and pushed away from the Sun, was
projected to the west in January, matching the direction of the
observed tail. By the time of the February observations, the Earth had
moved to the point that we were looking almost straight down this
tail. Unfortunately, the lack of UVOT \vband\ observations preclude
the comparison of larger dust grains between the two dates.

\subsubsection{Solar wind interaction in X-rays}\label{sec:luli_morx}
As discussed in the previous section, the OH distribution in the coma
was very symmetric and homogeneous during our observations. Past
modelling has indeed indicated a transition between thick and thin
regimes between $10^{28}$ and $10^{29}$\,\mps. Lulin's large gas
production rate (6--8\,\moleps, Sec.~\ref{sec:luli_gasp}) made it
collisionally thick to charge exchange by solar wind ions
(c.f. \citealt{lisse2005} and \citealt{bodewits2007}). It is therefore
to be expected that when viewed via X-ray emission, the comet would
appear as a crescent when observed at phase angles of around 90
degrees, with a significant sunward offset between the maximum of the
observed X-ray emission and the nucleus of the comet (cf. Hyakutake
\citep{lisse1996}; Linear S4, \citep{lisse2001}; C/2001 WM1
\citep{wegmann2005}). From the models by \citealt{bodewits2007}, most
charge exchange reactions in the coma of a comet with a gas production
rate of $\rm \sim\,10^{29}\,$\mps\ would occur $10^{3}$ - $10^{4}$\,km
from the nucleus (at a distance of 1\,AU from the Sun).

In Figure~\ref{fig:luli_imag} (lower panel) we plot X-ray contours
over a UVOT UV image of the comet (from the first observation, 31332
using filter \uvotw). As this X-ray image is constructed from images
taken at different Sun-Earth-comet angles and geocentric distances
(resulting in differing angular extents of the expected X-ray
emission), caution should be exercised when interpreting the implied
morphology. A very rough crescent shape can be seen in
Figure~\ref{fig:luli_imag} (lower panel). Assuming the
exposure-weighted distance of 0.685\,AU to the comet we find that the
X-ray maximum is offset in the sunward direction from the UV maximum
by $\sim$35,000\,km (foreshortened by the observing geometry). While
in line with other comet observations \citep{wegmann2005,
  dennerl2003}, this stand off distance is greater than expected by
the model results of \citet{bodewits2007}, who predict the brightness
peak around 10,000 km for Q = 10$^{29}$ molecules/s at a heliocentric
distance 1 AU, and less than predicted by the model of
\citet{wegmann2004}, who predict 66,000 km under the same
circumstances.
  

\subsection{Temporal variations}\label{sec:luli_temp}
Variations in the comet's X-ray luminosity are driven by variations in
the comet's gas production rate, in the solar wind ion flux, and by
the solar wind ion content. Along with its X-ray observations, \swift\
simultaneously obtained UV/optical observations of the comet, which
provide a measure of the comet's gaseous activity. The solar wind is
currently sampled by various spacecraft and data from these spacecraft
are the best proxies for near-Earth comet environments. In this
section we interpret our observations in terms of the variability of
both the cometary neutral gas content and the solar wind ion flux.

\subsubsection{Gas production rates}\label{sec:luli_gasp}
At the wavelengths covered by UVOT, comets are seen in sunlight
reflected by cometary dust, with several bright molecular emission
bands superposed. The \uvotw\ filter is well placed to observe the
three very strong OH vibrational transitions between 280 and
312\,nm. Lulin's spectrum was only slightly reddened ($<10$\% per 1000
\AA; \citealt{btest2011,lin2007}). We therefore assumed an un-reddened
solar continuum, and convolved this with the \uvotw\ and \vband\
filter transmissions to determine how much the continuum contributes
to the \uvotw\ flux. If we further assume that the flux in the \vband\
filter is dominated by continuum emission (while in truth it is
contaminated by the fluorescent emission of various molecules,
predominantly C$_2$ and NH$_2$). The OH flux is then given by
\textbf{Equation~\ref{eqn:luli_flux}}:

\begin{equation}
  F_{OH} = F_{\uvotw} - (0.15 \times F_{V})
  \label{eqn:luli_flux}
\end{equation}

\noindent where $F_{\uvotw}$ and $F_{V}$ are the fluxes measured in
the \uvotw\ and \vband\ filters. The factor 0.15 originates from the
ratio $F_{\uvotw}$/$F_{V}$ for a solar spectrum. From Table 1, on
\febd\ the continuum contribution (here assumed to be any
contributions other than from OH) to the flux measured in the \uvotw\
band was approximately 21\%; on \mard\ it was 17\%. We consider the
possibility that there is a non-negligible contribution from the
fluorescence emission of C$_2$ to the \vband\ measured flux. Using the
total C$_2$ flux within this band determined from our work with the
\swift\ grism \citep{btest2011}, we estimate the contribution to be
0.25\, \uvfu. This corresponds to an over-subtraction of 8\% in the
calculation of the OH flux using Equation~\ref{eqn:luli_flux}. As this
is well within the errors quoted in Table 1, we consider this
contamination of marginal consequence.

Using the molecules' fluorescent rate \citep{schleicher1988}, the
number of OH molecules in the aperture can be derived
(\citealt{feldman2005}). The results are listed in Table 1. To derive
production rates, we modelled the OH and water distribution, and
compared this with the measured OH number densities. This model is
based on the Haser model \citep{haser1957, festou1981} and is
discussed in greater detail in \citet{btest2011}.

The use of broad band filters implies that our results are a
relatively crude estimate of the comet's water production and we
estimate the systematic error to be around 25\%. The resulting OH
production rates are summarised in Table 1.  No \vband\ observations
where obtained on \jand, and without accounting for the continuum
removal we find a 3-sigma upper limit of $\rm
Q_{OH}$\,$<$\,1\,\molepsb, in good agreement with the production rates
of 5.8$\pm$0.6 (1-sigma) to 6.9$\pm$0.6\,\moleps\ measured with the
grism \citep{btest2011}. On \febd\ and \mard\ the OH production rates
were 5.5$\pm$1.0\,\moleps\ and 4.2$\pm$1.0\,\moleps\textbf{,}
respectively.

\begin{figure}
  \centering
  \includegraphics[width=0.48\textwidth, bb=20 8 360 315, clip]{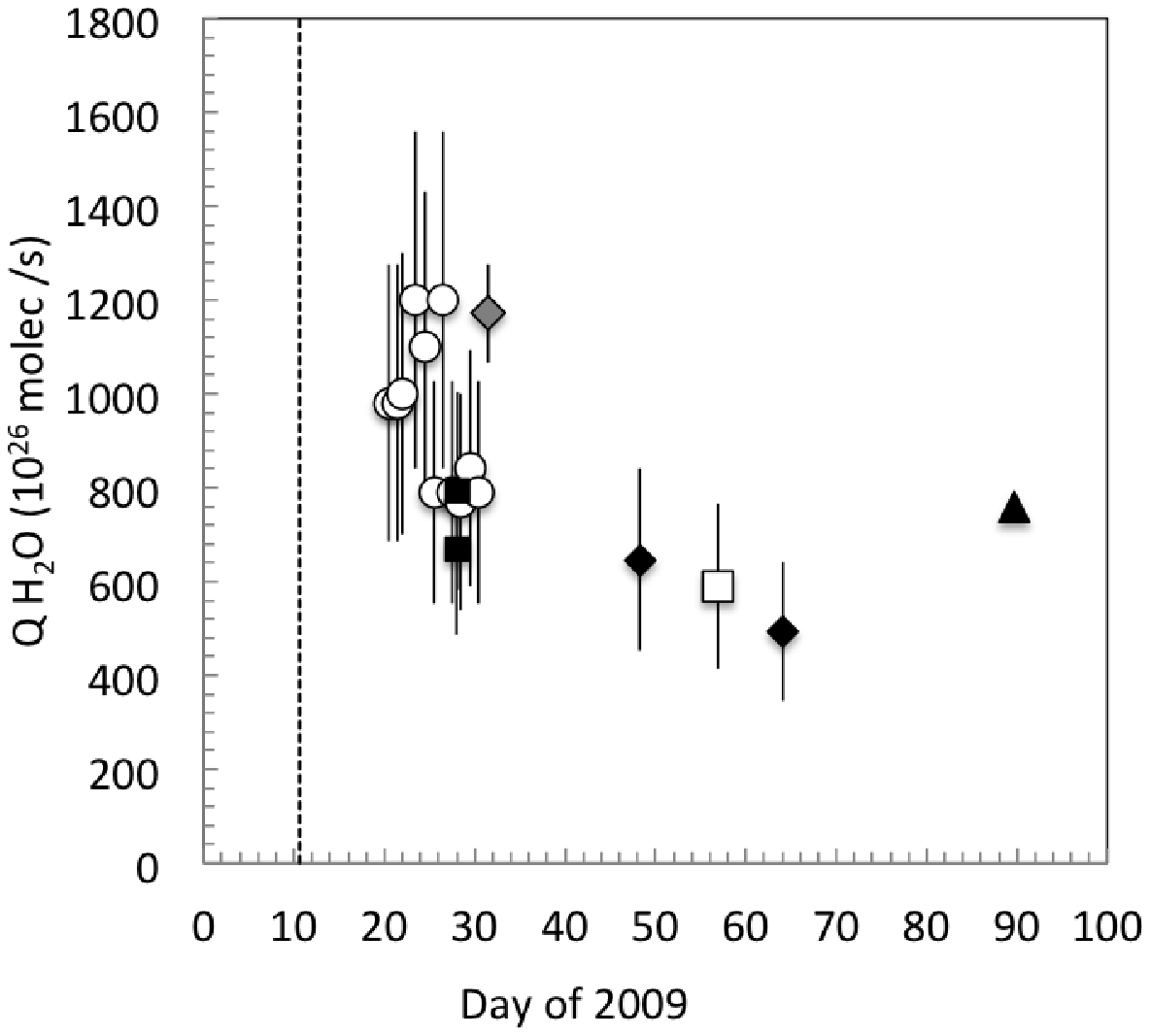}
  \caption{Gas production rates of \water\ vs time, calculated for our
    observing periods on \febd\ and \mard\ (solid, black diamonds). We
    also plot \water\ production rates from \citealt{combi2009}
    (circles), Schleicher 2009 (private communication, open squares),
    \citealt{btest2011} (filled squares), \citealt{ootsubo2010}
    (filled triangle) and \citealt{bonev2009} (solid, grey
    diamond). The dashed vertical line indicates the date of the
    perihelion passage of the comet.}
  \label{fig:luli_watt}
\end{figure}

The OH production rates were converted to water production rates
assuming a branching ratio for the formation of OH from \water\ of
85\% \citep{huebner1992,combi2004}. In Figure~\ref{fig:luli_watt} we
show these rates versus time, for this work and other measurements
(\citealt{combi2009, bonev2009, ootsubo2010, btest2011}, and
D. Schleicher (private communication). Our measurements indicate that
the comet's activity was slowly decreasing after it had reached
perihelion on UT Jan. 10.64, 2009, in good agreement with other
measurements.

Several studies found that gas production rates varied on a day-to-day
basis by as much as 50\%, suggesting strong diurnal effects on the
comet \citep{btest2011, combi2009, knight2009}. Based on the phasing
of CN jets, \citet{knight2009} found a rotation period for comet Lulin
of 42$\pm$0.5\,hr. Even though our UVOT broadband observations on
\febd\ and \mard\ sampled the comet every hour for about half a day
and thus we did not observe the comet for a full rotation period, our
data showed no significant short term flux variations.


To estimate the comet's neutral gas contribution to its X-ray
variability one needs to know the number of neutral coma molecules in
the FOV. It can be assumed that the X-rays are mostly driven by SWCX
with H$_2$O and OH \citep{bodewits2006b}. We therefore need to know
the sum of the number of H$_2$O and OH molecules in the aperture. If
the aperture were large enough, the total number of OH molecules would
be 0.85 times the number of H$_2$O molecules (the branching ratio of
the H$_2$O to OH photodissociation process), and the total number of
neutral molecules in the FOV would thus be 2.2 times the number of OH
molecules present. However, in apertures smaller or equivalent to the
Haser scale-lengths for \water\ to OH dissociation (of the order of
$10^5$\,km at 1\,AU from the Sun), the number of H$_2$O molecules will
be larger than the number of OH molecules. For the apertures used here
($\sim 3.1 \times 10^{5}$\,km, $\sim 1.45\times 10^{5}$\,km and $\sim
1.51\times 10^{5}$\,km radius, for the three observing periods), the
total number of \water\ plus OH molecules, calculated by Haser
modelling, are 1.54, 1.56, and 1.57 times the number of observed OH
molecules on \jand, \febd, and \mard, respectively. This factor
incorporates the lifetimes of the H$_2$O and OH molecules with
cometary distance from the Sun and accounts for the proportion of
H$_{2}$O compared to OH observed depending on aperture size. For the
\jand\ measurement the aperture is much larger than the Haser scale
length, whereas the factors for \febd\ and \mard\ are adjusted to
account for the smaller apertures used.

To estimate the number of water and OH molecules on \jand, when we did
not have any \vband\ images to subtract the continuum emission from the
\uvotw\ filter, we used the OH production rates derived from the grism
observations and used our Haser model to estimate the number of
molecules in a 400 arcseconds radius aperture.

\subsubsection{Solar wind}\label{sec:luli_swda}
\begin{figure}
  \centering
  \includegraphics[width=0.475\textwidth, bb=18 10 670 485, clip]{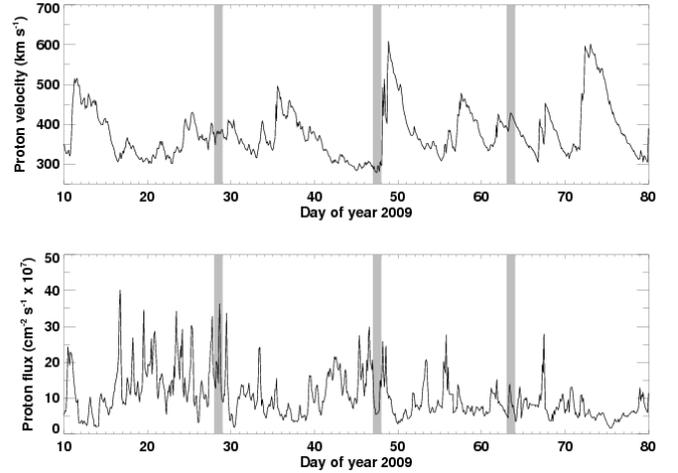}
  \caption{Solar wind parameters, taken from \stereoa, that have been
    co-rotationally mapped to the location of the comet. Upper panel:
    proton velocity. Bottom panel: proton flux. The three observing
    periods are highlighted within the light-grey shaded areas.}
  \label{fig:luli_swlt}
\end{figure}

We took data from \stereoa\ to investigate the behaviour of the solar
wind in early 2009, shown in Figure~\ref{fig:luli_swlt}. We plot the
solar wind proton speed and flux, mapped to the approximate location
of the comet. \stereoa\ is found ahead of the Earth in its orbit on
the ecliptic plane. In order to map the solar wind from \stereoa\ to
the position of the comet at the three observing periods, we used the
time shift procedure described by \citet{lisse1999} and
\citet{neugebauer2000}. The calculations are based on the comet
ephemeris, the location of the comet and \stereoa, and the bulk solar
wind velocity measured by \stereoa. With this procedure, the time
delay between an element of the co-rotating solar wind arriving at the
proton monitor and the comet can be predicted. Although this mapping
is an approximation that does not take propagating shocks or
latitudinal structures in the wind into account \citep{neugebauer2000,
  bodewits2007}, the relatively close proximity of the comet and its
low heliocentric latitude ($<$1 degree during our observations) aid
its reliability.

On \jand, there is a delay of 2.5 days between the solar wind measured
by \stereoa\ and the solar wind arrival at the comet, mostly due to
the longitudinal difference. On \febd, this delay is 0.8 days, and on
\mard\ the comet leads \stereoa\ by 1.8 days. It is of note that the
uncertainty in the solar wind arrival time is the largest in January,
when the comet was still at 1\,AU from Earth versus 0.5\,AU in
March. 

The averaged solar wind proton fluxes, were 15, 5 and 14 $\times
10^7$\,\cmpers\ for \jand, \febd, and \mard\ respectively. The average
solar wind proton speeds for the three observing periods were $\sim$
370, 280 and 420\,\kms, implying solar wind densities of 4.2, 1.9 and
3.4\,\cmd. Inspection of long term solar wind data indicate the
presence of several CIRs (a compression region followed by a
rarefaction, as described in Section~\ref{sec:luli_intr}) during the
early part of 2009. The most notably of these occurred just after the
second observing period, as indicated by sharp increases in solar wind
proton velocity. In addition a weaker CIR occurred during the third
observing period, shown in the upper panel of
Figure~\ref{fig:luli_swlt}. The low solar wind proton densities may be
explained by the presence of these CIRs. The solar wind data further
show that a CIR reached the comet around \febf\ (day of year
36). Considering the uncertainty in the time delay of our propagation
model, we consider it likely that it was this CIR that caused a
disconnection event at the comet, reported by \citealt{shi2011}.

\subsubsection{X-ray lightcurve}\label{sec:luli_xrlc}
\begin{figure}
  \centering
  \includegraphics[width=0.475\textwidth, bb=0 55 465 820, clip]{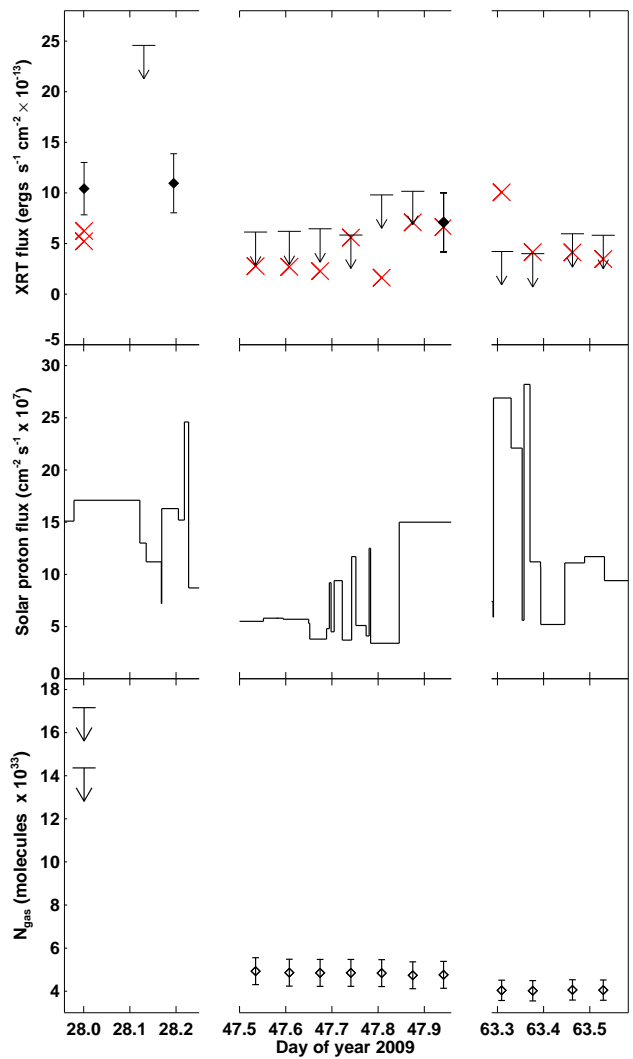}
  \caption{Upper panel: XRT \bone\ to \btwo\,keV background-corrected
    fluxes (diamonds) or upper limits (downward arrows) where
    appropriate, plotted against a time-axis given in the day of
    year. We also plot the predicted X-ray flux (red crosses,
    normalised to the tenth observation) as described in the
    text. Middle panel: solar wind proton flux as measured by
    \stereoa\ for early 2009. Lower panel: $N_{gas}$ (open diamonds),
    or upper limits (downward arrows) where appropriate.}
  \label{fig:luli_swpf}
\end{figure}

In Figure~\ref{fig:luli_swpf} (upper panel) we plot the
background-corrected XRT fluxes in the energy band \bone\ to
\btwo\,keV for each observation (accounting for the areas of the
spectral extraction regions to convert to the stated flux units), or
the upper limit where appropriate. We also show the predicted X-ray
flux, calculated from the product of solar wind proton flux and the
number of neutral molecules in the FOV, corrected for the Earth-comet
distance (Equation~\ref{eqn:luli_pred}):

\begin{equation} 
  F_x  = k * \frac{N_{\rm gas}*F_{\rm sw}(r_{\rm h})}{\Delta^{2}}
  \label{eqn:luli_pred}
\end{equation}

\noindent where $k$ is a normalisation constant, $N_{\rm gas}$ is the
total number of water group molecules available to contribute to the
charge exchange process derived from our UVOT observations (see
Section~\ref{sec:luli_gasp}), $F_{\rm SW}(r_{\rm h})$ are the solar
wind proton fluxes at the comet (as described in
Section~\ref{sec:luli_swda}), and $\Delta$ are the Earth-comet
distances (as listed in Table 1). We normalised to the flux measured
in the tenth XRT observation, as this data point had both a
constrained X-ray and UVOT measured flux. In
Figure~\ref{fig:luli_swpf} (middle panel) we plot the solar wind
proton flux, at the three observing periods as described below. The
lower panel of Figure~\ref{fig:luli_swpf} shows $N_{\rm gas}$ over the
periods of the observations.

There are several competing processes that contribute to the predicted
X-ray flux as given in Equation \ref{eqn:luli_pred}. The gas
production rate was 50\% larger in January compared to the other
observing periods, however the comet was at half the distance from
Earth in February and March. The average proton flux was comparable
during \jand\ and \mard, and about 40\% lower during our \febd\
observations. The contributing parameters to the predicted lightcurve
compete to cancel each other out, resulting in a relatively flat
predicted X-ray lightcurve. Thus our measurements do not exhibit any
large variations, and our predicted and measured lightcurves are in
reasonably good agreement.

The solar proton flux is only a proxy for the heavy ion flux of the
solar wind and this heavy ion content may vary considerably due to
variations in either solar wind atomic abundance or ion temperature
and charge state. Thus we looked at additional available solar wind
\opluss\ flux data during the time of our observations (e.g. from the
Advanced Composition Explored (\ace), \citealt{stone1998}), located at
Lagrangian Point L1). Unfortunately the available data were sparse and
of low quality.

Had the comet interacted with a ICME (a large cloud of plasma ejected
from the Sun, as described in Section~\ref{sec:luli_intr}) during our
observations, we would have in general expected a large increase in
the measured X-ray flux and spectral hardness due to the presence of
highly charged ions (as shown by the examples of an ICME interacting
with the exosphere of the Earth; \citealt{ezoe2011,
  carter2010}). Approximately 10\% of ICMEs appear to exhibit only
weak compositional anomalies \citep{richardson2004}. The long-term
trend data of Figure~\ref{fig:luli_swlt} suggested that during at
least one, if not two of the observing periods, the comet had sampled
a CIR. CIRs occur when a fast solar wind stream piles up against a
slow solar wind stream. CIRs show average compositional signatures
similar to average fast and cool solar wind \citep{mason2008}. Hence,
the comet may have sampled, or partially sampled, a fast, cool plasma
that would have exhibited lower abundances of highly-charged heavy
ions that are responsible for X-ray emitting, charge-exchange
interactions. The X-ray spectrum of Lulin is indicative of a higher
solar wind charge state than the very soft X-ray spectrum of comet
17P/Holmes \citep{christian2010}, which showed no X-ray emission above
0.4\,keV and was postulated to be the result of an interaction with
the cool, polar solar wind.

\section{Conclusions}\label{sec:luli_conc}
We have used the unique capabilities of the \swift\ observatory to
simultaneously observe comet C/2007 N3 (Lulin) via UV and X-ray
emission. We have carefully transformed X-ray events from 14 \swift\
XRT pointings towards comet Lulin to produce the optimum
exposure-corrected X-ray image. We have also analysed \swift\ UVOT
exposures when available and used this data to observe variations in
the morphology of the comet in particular with regards to the dust and
ion tails. We see the peak of the cometary X-ray emission offset from
the peak in the UV emission, and this offset is displaced towards the
Sun, indicating that the coma was collisionally thick to charge
exchange from the nucleus out to about 35,000\,km, as expected for a
comet with a gas production rate of $\rm Q_{H_{2}O}$ = 6\,\moleps.


A model applied to a background-corrected X-ray spectrum was
consistent with emission from charge exchange, although individual
lines were heavily blended. The spectrum, only discernible below
1\,keV, also suggested that the comet had sampled a cool solar wind,
with less highly ionised oxygen than a quiescent (warm) equatorial
solar wind. Such a spectrum may be expected if the comet had
encountered a stream of fast wind originating at the cold base of the
corona and flowing through a coronal hole. We have used the solar wind
proton flux as a proxy for solar wind activity throughout this work,
as investigations regarding solar wind minor ion fluxes were hampered
by a lack of high cadence data from spacecraft solar wind ion monitors
during the \swift\ observations. What heavy ion data trends that do
exist suggest there were indeed numerous CIRs during the period of the
\swift\ Lulin observations.

We combined our simultaneous X-ray and UV observations with measured
solar wind data to explain the measured X-ray lightcurve. We use data
from the \swift\ UVOT, employing the \uvotw\ and \vband\ filters to
measure the number of OH molecules in the FOV and estimate the number
of water molecules released by the comet. We used the water production
rates of the comet and solar wind data mapped to the location of the
comet to estimate the expected X-ray flux. We demonstrate that Lulin's
X-ray brightness was determined by the comet's activity and the
observing geometry, as well as variations in the solar wind.

\begin{acknowledgements} The authors would like to thank the \swift\
  instrument teams for all their help and advice, and T. L. Farnham
  (University of Maryland) for a most helpful discussion on the
  observing geometry and for running his synchrone/syndine calculation
  for these observations. The authors also thank S. Lepri (University
  of Michigan) for the interpretation of the space weather
  data. J. A. Carter and A. M. Read gratefully acknowledge funding by
  the Science and Technology Facilities Council, U.K. We thank the
  anonymous referee for the suggestions for improvement which have
  greatly enhanced the manuscript.
\end{acknowledgements}

\bibliographystyle{aa} 
\bibliography{phd_gen_aa} 

\end{document}